\documentclass{article}
\usepackage[utf8]{inputenc}
\usepackage{a4wide}
\usepackage{xcolor}
\usepackage{amsmath}
\usepackage{amssymb}
\usepackage{graphicx}
\usepackage{hyperref}
\usepackage{multirow}
\usepackage{array}

\definecolor{darkgreen}{rgb}{0,0.4,0}

\newcommand{\GeV}{\,\text{GeV}}

\newcommand{\OXaff}{Rudolf Peierls Centre for Theoretical Physics, \\
  \normalsize
  Clarendon Laboratory, Parks Road,
  University of Oxford, Oxford OX1 3PU, UK}
\newcommand{\ASCaff}{All Souls College, Oxford OX1 4AL, UK}
\newcommand{\Chiaff}{University of Chicago, 5640 South Ellis Ave. Chicago, IL 60637}
\newcommand{\MPIaff}{Max Planck Institute for Physics, F\"ohringer Ring 6, 80805 Munich, Germany}
\newcommand{\TUMaff}{Physik-Department, Technische Universit\"at M\"unchen, James-Franck-Strasse 1, 85748 Garching, Germany}
\newcommand{\GZemail}{zanderi@mpp.mpg.de}

\title{The Higgs boson Turns Ten}
\author{Gavin P. Salam,$^{1,2}$ Lian-Tao Wang,$^{3}$ Giulia Zanderighi$^{4,5,6}$\\
    \normalsize $^1$ \OXaff,\\
    \normalsize $^2$ \ASCaff,\\
    \normalsize $^3$ \Chiaff,\\
    \normalsize $^4$ \MPIaff\\
    \normalsize $^5$ \TUMaff\\
    \normalsize $^6$ \GZemail\\    
}

\date{}

\begin{document}

\maketitle

\begin{abstract}
  The discovery of the Higgs boson, ten years ago, was a milestone
  that opened the door to the study of a new sector of fundamental
  physical interactions.
  We review the role of the Higgs field in the Standard Model of
  particle physics and explain its impact on the world around us.
  We summarize the insights into Higgs physics revealed so far by ten years of work, discuss what remains
  to be determined, and
  outline potential connections of the Higgs sector with unsolved
  mysteries of particle physics.
\end{abstract}
\bigskip

Ten years ago, on the 4$^\text{th}$ of July 2012, scientists and journalists
gathered at CERN, and remotely around the world, for the announcement
of the discovery of a new fundamental particle, the Higgs boson.
The discovery, by the ATLAS~\cite{ATLAS:2012yve} and
CMS~\cite{CMS:2012qbp} collaborations at the Large Hadron Collider
(LHC), came almost fifty years after theorists had postulated the existence
of such a particle.
The significance of the discovery was not only that a new, long
awaited particle had been found, but that the existence of this
particle provides first direct evidence that surrounding us there is a
new kind of fundamental ``field'', known as the Higgs field.

Fields in physics are familiar in everyday life, for example 
in the form of the earth's magnetic field, and its
impact on the needle of a compass.
The most important difference between the Higgs field and a magnetic field is that  if one removes the magnetic source,
the magnetic field disappears.
In contrast, the Higgs field is non-zero everywhere, all the time, independently of whether
anything else is present in the universe.
In a way, it is reminiscent of the ancient Greek concept of Aether
with the crucial difference that it is consistent with
Einstein's theory of special relativity.

Physicists' current theory of fundamental particles and forces is
known as the Standard Model, a theoretical framework that provides a
description of elementary particles and the forces that make them
interact with one another, with the exception of gravity. Within the
Standard Model, the Higgs field is essential to describe the world as
we know it.

\begin{figure}[t]
  \centering
   \includegraphics[width=\textwidth,page=1]{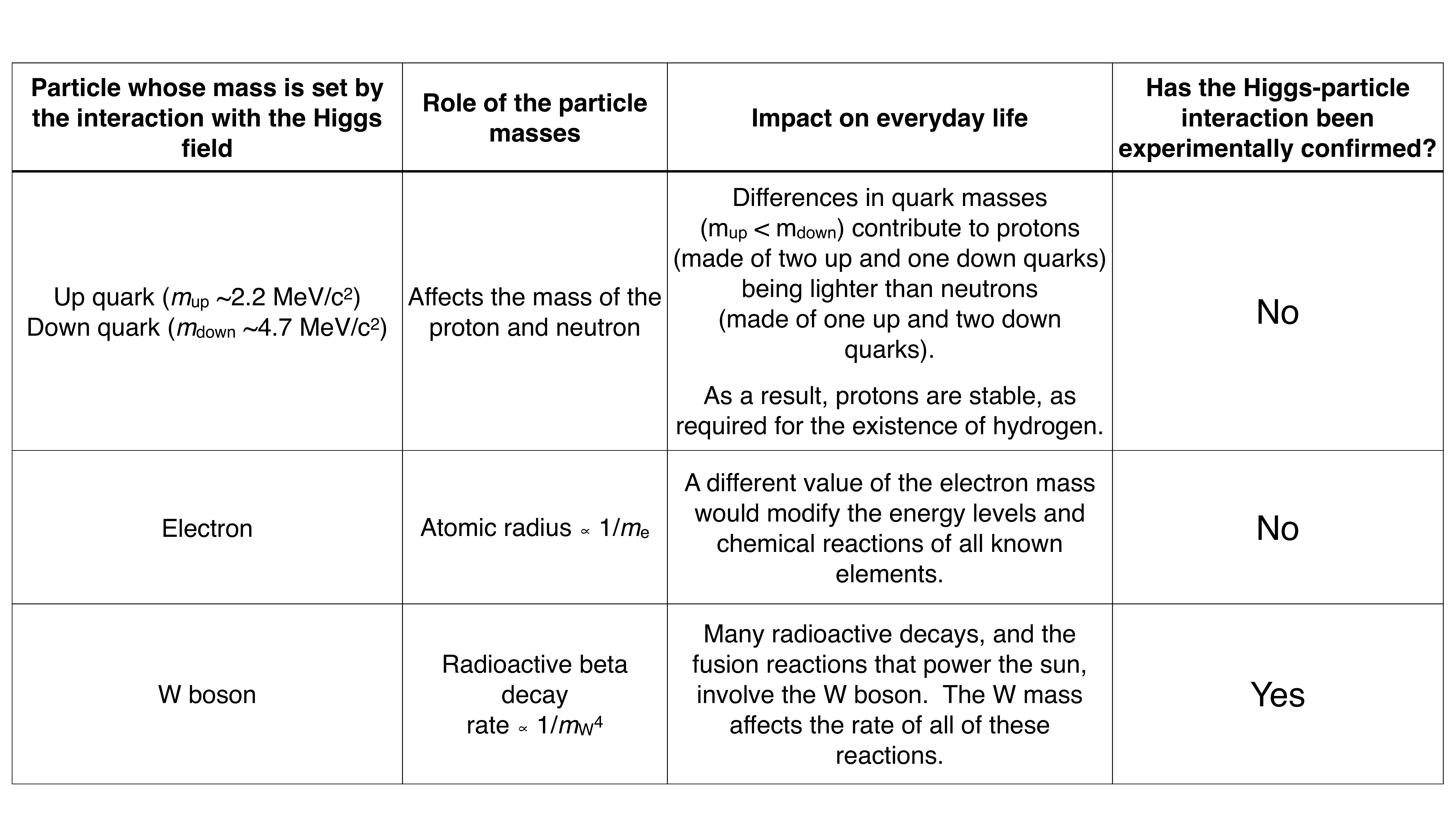}
  \caption{{\bf Ways in which the Higgs boson affects the world around us.}
    Three examples of how particle masses~\cite{ParticleDataGroup:2020ssz} play a crucial role in determining
    the physical nature of the world in which we live.
    In all three cases, the Standard Model suggests that the
    corresponding particle
    masses arise from interactions of those particles with the Higgs
    field. 
    The last column indicates whether or not we have clear
    experimental indications that confirm that hypothesis.
  }  
  \label{fig:mass-impact}
\end{figure}

As we shall see below, the strength of the interaction between any
particle and the Higgs field directly affects a fundamental property
of that particle: its mass.
As such~\cite{Quigg:2009xr},
it ultimately determines the size of atoms, makes the proton stable
and sets the timescale of radioactive ($\beta$) decays, which for example
impacts the lifetime of stars, cf.\ Fig.~\ref{fig:mass-impact}.
Yet, in everyday life, we do not notice that the Higgs field is
all around us.
The only way we have of revealing the Higgs field is to perturb it, a
little like throwing a stone into water and seeing the ripples.
The particle known as the Higgs boson is the manifestation of such a
perturbation. 

The significance of its discovery in 2012 
was such that 
the Nobel prize was awarded one year later to François Englert
and Peter Higgs who, with the late Robert Brout, were the first to
discuss the potential importance of such a field for fundamental
physics~\cite{Englert:1964et,Higgs:1964ia,Higgs:1964pj}.
Since then, the Higgs boson has become a powerful tool to study the
ways in which the underlying Higgs field affects the fundamental
particles of the Standard Model.
Furthermore, the ubiquity of the Higgs field means that the Higgs
boson is, today, widely used in the search for signatures of particles or
effects that are hitherto unknown and lie outside the Standard Model.

\section*{The Higgs boson in the Standard Model}

In the Standard Model, aside from the Higgs boson, there are two kinds
of particles.
There are fermions, such as the up and down quarks and the electron,
which make up ordinary matter.
These specific particles (together with one of the three neutrinos) are
called first generation fermions.
Two further sets of fermions (second and third generations), involve
heavier particles, not normally present in the world around us.
Additionally, there are the force-carriers: the photon, the $W$ and
$Z$ bosons and the gluon, collectively called vector bosons.
When these are exchanged between two fermions, they create an attractive or repulsive  force
between those fermions: photons carry the electromagnetic force, $W$
and $Z$ bosons the weak force and gluons the strong force.

In the 1960s, as physicists were taking the first steps towards assembling this picture, it remained unclear whether a self-consistent theory that included massive force carriers could be constructed.
This question was being posed in the context of nuclear physics as well as superconductivity in condensed matter physics. Researchers found that such a theory was ultimately
possible if one introduced an interaction of the force carriers with
a ``Higgs'' field, and if one could also engineer a non-zero value for
that
field~\cite{Anderson:1963pc,Higgs:1964ia,Higgs:1964pj,Englert:1964et,Guralnik:1964eu,Kibble:1967sv}.

As the electroweak part of the Standard Model was being
developed~\cite{Glashow:1961tr,Salam:1964ry,Weinberg:1967tq}, 
interactions of particles with a Higgs field were to become a central part of 
its formulation, especially in order to generate masses for the $W$
and $Z$ bosons, as required for consistency with experimental
observations, while photons and gluons remain massless.

Remarkably, interactions with the Higgs field also provided a consistent
theoretical mechanism for producing fermion masses:
each fermion interacts with the Higgs field with a different strength
(or ``coupling''), and the stronger the interaction, the larger the
resulting mass for the particle.
Within the Standard Model
the interaction is known as a ``Yukawa''
interaction~\cite{Yukawa:1935xg}.
Thus any question about the origin of the masses of fermions reduces
to a question about the origin of the fermions' interactions with the
Higgs field.

\begin{figure}[t]
  \centering
  \includegraphics[width=0.55\textwidth]{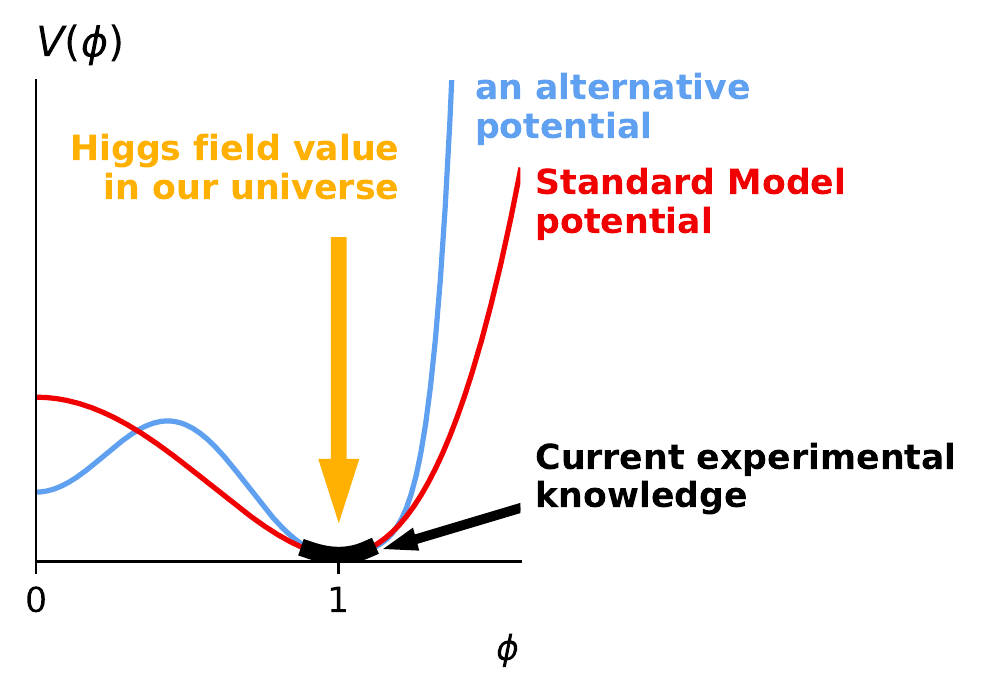}
  \caption{{\bf Higgs potential}.
    Potential energy density $V(\phi)$ associated with the
    Higgs field $\phi$, as a function of the value of $\phi$.
    The red curve shows the potential within the Standard Model.
    The Higgs field has a value corresponding to a minimum of the
    potential and the region highlighted in black represents our current
    experimental knowledge of the potential.
    Alternative potentials that differ substantially from the Standard
    Model away from that
    minimum ({\it e.g.}\ the blue curve) would be equally consistent with
    current data.  
}
  \label{fig:higgs-potential}
\end{figure}

Why is the Higgs field non-zero in the first place?
According to the Standard Model there is a
potential energy density associated with the value of the Higgs field
and the lowest potential energy corresponds to a non-zero value
of the Higgs field.
The Standard Model potential has a form dictated by 
internal consistency conditions.
With some simplifications, labeling the magnitude of the Higgs field
as $\phi$, the potential has the form
\begin{equation}
  \label{eq:Vphi}
  V(\phi) \propto -\phi^2 + \frac12 \phi^4\,.
\end{equation}
This is illustrated by the red line in
Fig.~\ref{fig:higgs-potential}.
The minimum of the potential, {\it i.e.}\ the energetically most favourable
choice for $\phi$, lies at a value of $\phi$ that is non-zero,
$\phi = 1$.
An important implication of the Higgs field's non-zero
constant value is the impossibility to carry angular momentum,
or more technically having ``spin 0''.
A non-zero value for the spin would break at least one of the well-tested
space-time symmetries.
Hence, the excitation of the Higgs field, the Higgs boson, must be a
spin-0 particle and is in fact the only known fundamental particle with this property.

One of the reasons for the central importance of the discovery of the
Higgs boson was that it finally made it possible to
start testing the remarkable theoretical picture outlined above.
It is not possible to probe the interactions of a given particle
with the Higgs field. However, one can instead measure a particle's
interaction with the excitations of the Higgs field, {\it i.e.}\ with a
Higgs boson.
If the Standard Model provides the correct picture for the generation
of mass, the strength of any particle's interaction with the Higgs
boson has to be directly related to that particle's mass.

Aside from providing a powerful way of testing the Higgs mechanism,
the interaction of the Higgs boson with other particles is intriguing
because it implies the existence of a ``fifth force'', mediated by the
exchange of Higgs bosons. 
The fact that such a force is stronger for heavier particles makes it
qualitatively different from all other interactions in the Standard
Model, whose interaction strengths come in multiples of some basic
unit of charge, like the electron charge for the electric force.
The pattern is, if anything, more reminiscent of gravity, but with
important differences.
One is that the force mediated by the Higgs boson is
active only at very short distances,
while Einstein's gravity 
acts over all distance scales.
Another is that the Higgs boson couples directly only to elementary
Standard Model particles.
In contrast gravity couples to the total mass.
In ordinary matter, that total mass is much larger than the sum of the
elementary particle masses, because the strong force contributes substantially to the proton and neutron masses~\cite{Durr:2008zz}.

\section*{What we know so far and how}

The Higgs mechanism provides the simplest model to explain particle
masses in a way that is consistent with the electroweak interactions.
As physicists we should seek to establish whether it is the model
chosen by nature.

Experimental studies of the Higgs boson take place at particle
colliders.
The likelihood of producing a Higgs boson in a collision becomes larger
when the particles that collide interact strongly with the Higgs
field, {\it i.e.}\ when they are heavy.
At the high centre-of-mass energies that are required, particle
physicists know how to collide just two things: protons and electrons,
as well as their anti-particles.
That poses an issue, because electrons and the particles that make up
protons are light, {\it i.e.}\ they interact only very weakly with the Higgs
boson.

\begin{figure}
  \centering
  \includegraphics[width=0.6\textwidth]{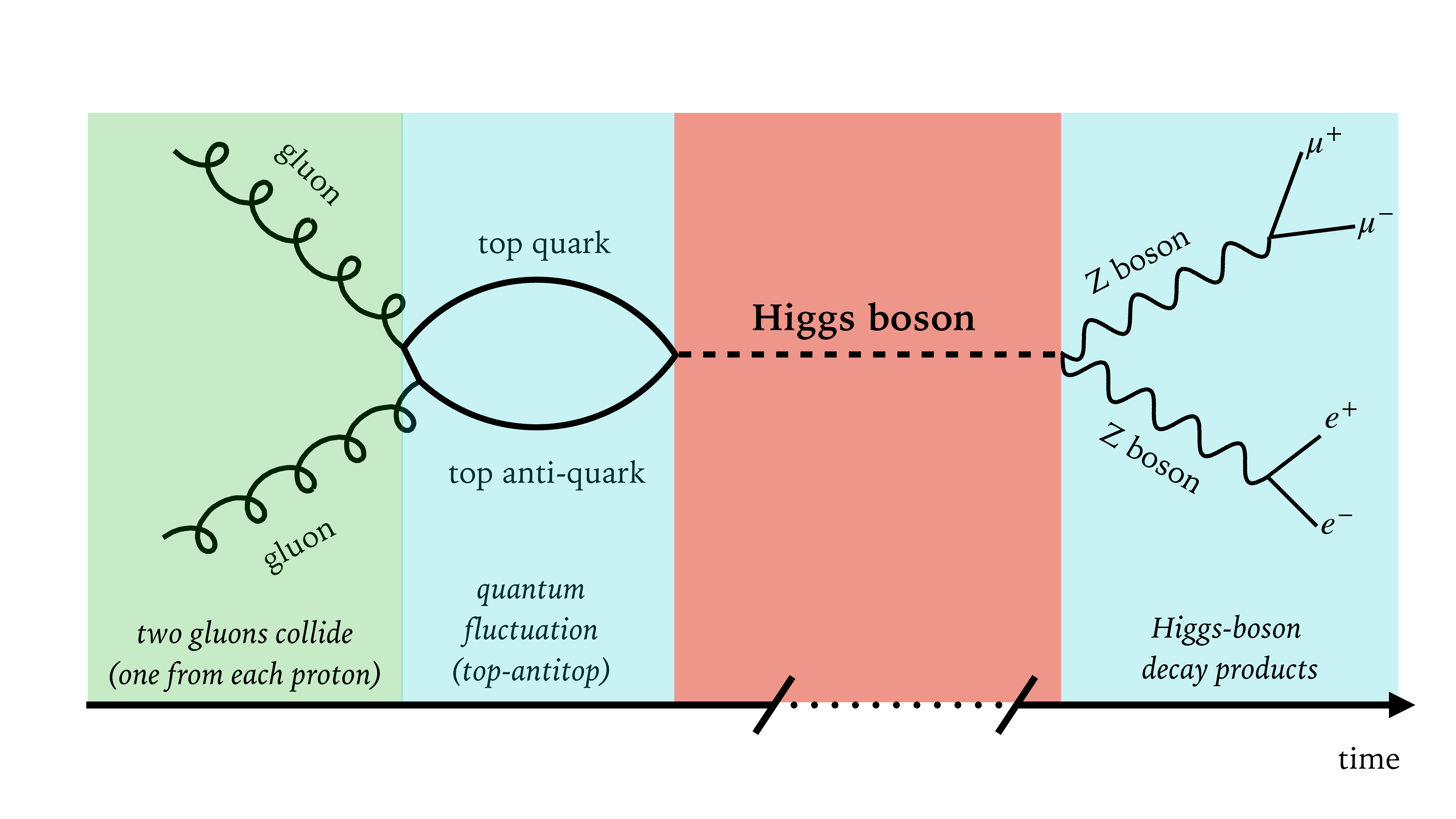}%
    \includegraphics[width=0.4\textwidth]{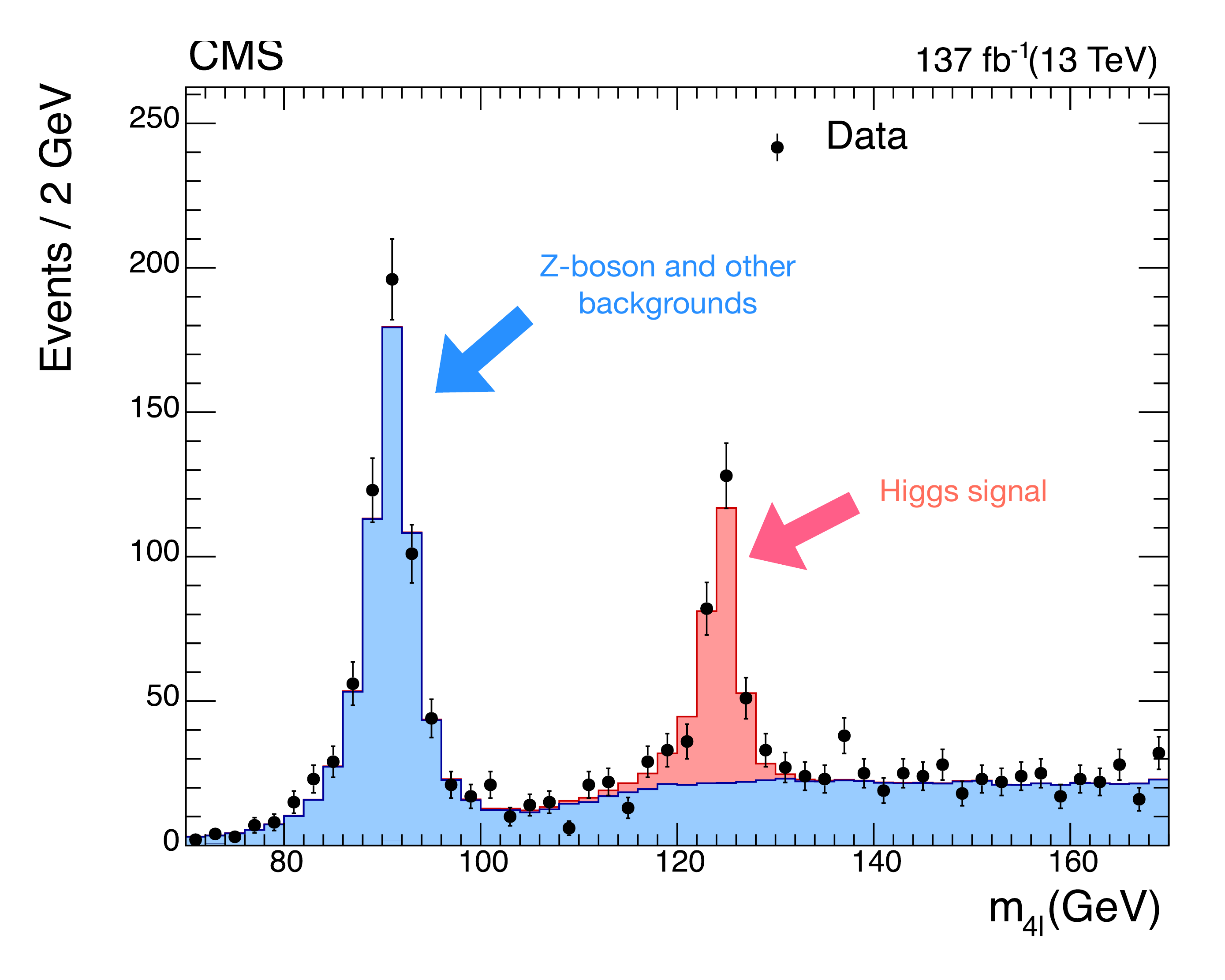}
  \caption{{\bf Higgs production at the LHC}. Left:
    Illustration of one process for the production and decay of a Higgs
    boson at the LHC.
    Right: total centre-of-mass energy of four leptons (electrons
    and/or muons and their anti-particles); the peak around
    $125\,\text{GeV}$ corresponds to decays of Higgs bosons, while the
    peak near $91.2\,\text{GeV}$ corresponds to decays of single $Z$
    bosons (not Higgs-induced), adapted from Ref.~\cite{CMS:2021ugl}.
    The decay to $Z$ bosons was one of the channels used
    for the Higgs boson discovery, with the other important
    discovery channels being the decay to two $W$ bosons and that to two
    photons (the latter proceeds via a quantum fluctuation with top
    quarks and $W$ bosons). 
  }
  \label{fig:higgs-prod-decay}
\end{figure}

The approach of particle physicists is to exploit the occasional production of heavy particles in the high-energy collision of light particles, and to then have 
those heavy particles produce a Higgs boson.
CERN's LHC collides protons, which are mostly made of up and down
quarks and gluons.
The most frequent way of producing a Higgs boson
is for a pair of gluons, one from each proton, to collide and
create a top quark and a top anti-quark as a very short-lived quantum
fluctuation.
The top-quark is the heaviest known particle (about 184 times the
proton mass) and so the top and anti-top quarks interact
strongly with the Higgs field, thereby occasionally producing a Higgs
boson.
A short while later (about $10^{-22}\text{s}$), the Higgs boson decays.
About $2.6\%$ of decays are to a pair of $Z$ bosons, which themselves
also decay almost immediately, for example each to an
electron-positron or muon anti-muon pair (so-called charged leptons),
a distinctive experimental signature.
This sequence is illustrated in Fig.~\ref{fig:higgs-prod-decay} (left).

The ATLAS and CMS experiments at the LHC select events with four such
leptons and record the total of the energy of the leptons (in their
centre-of-mass frame).
There are a variety of ways in which four leptons can be produced, but
for those events where they come from a Higgs-boson decay, the total
energy is expected to cluster around the Higgs mass --  
the red peak in Fig.~\ref{fig:higgs-prod-decay} (right).
That red peak provides considerable information:
(1) the existence of the peak near $125\GeV$ tells us that there is a
new particle, the Higgs boson;
(2) the position of the peak indicates the Higgs boson mass;
(3) other features of the events in the peak, for example the relative
angular distributions of the leptons (not shown in the figure),
confirm that the Higgs boson carries no intrinsic angular momentum,
{\it i.e.}\ it is a spin-0 particle;
(4) the number of events in the peak is sensitive to the interaction
strength of the Higgs boson both with top quarks and with $Z$-bosons.
This last point is crucial
because the Standard Model Higgs mechanism
predicts a very specific
interaction strength of each particle with the Higgs boson.
Point (4) provides us with a first test of this hypothesis.

\begin{figure}
  \phantom{x}\hspace{-0.5cm}
  \includegraphics[width=0.65\textwidth]{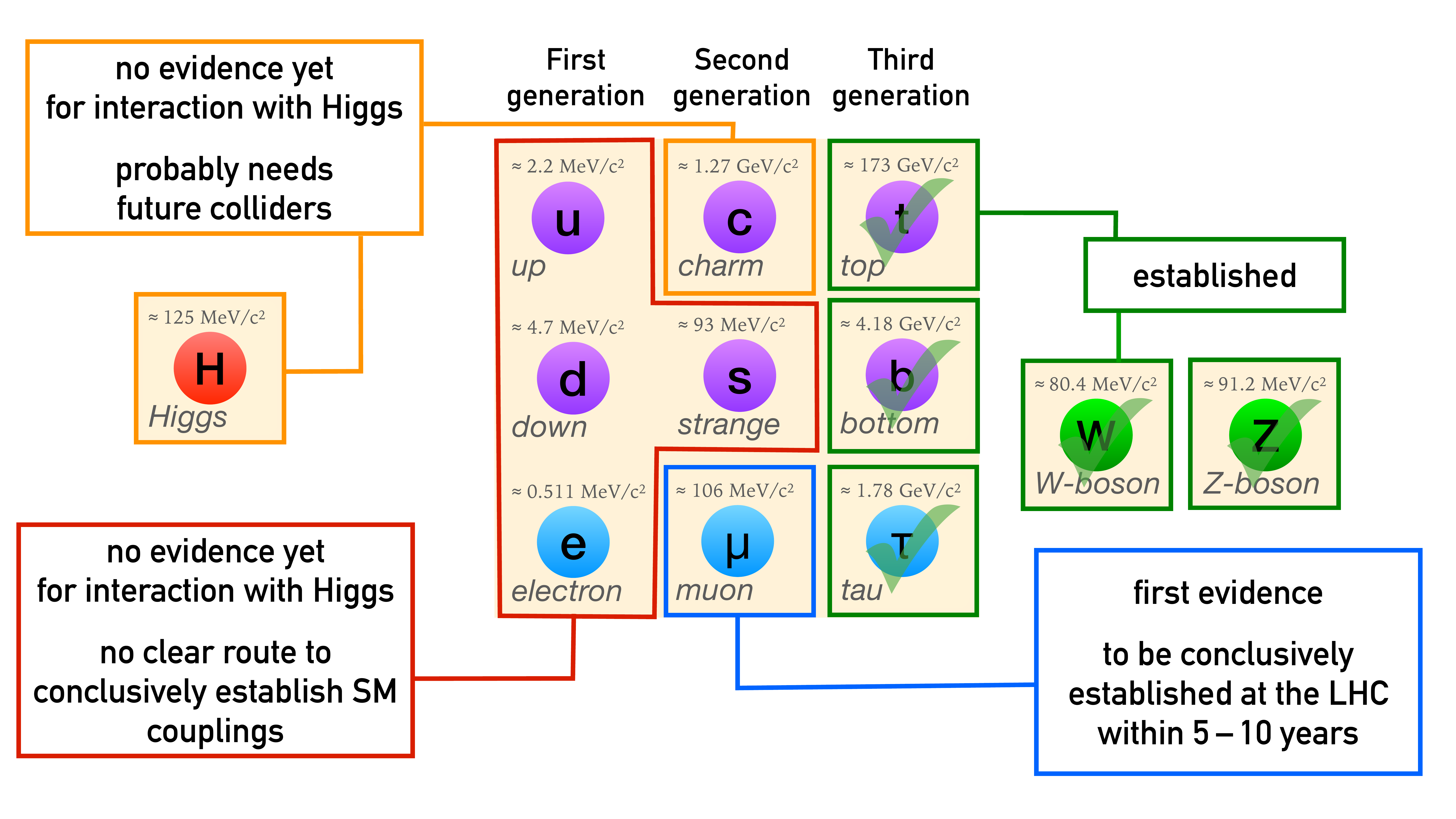}%
  \includegraphics[width=0.36\textwidth]{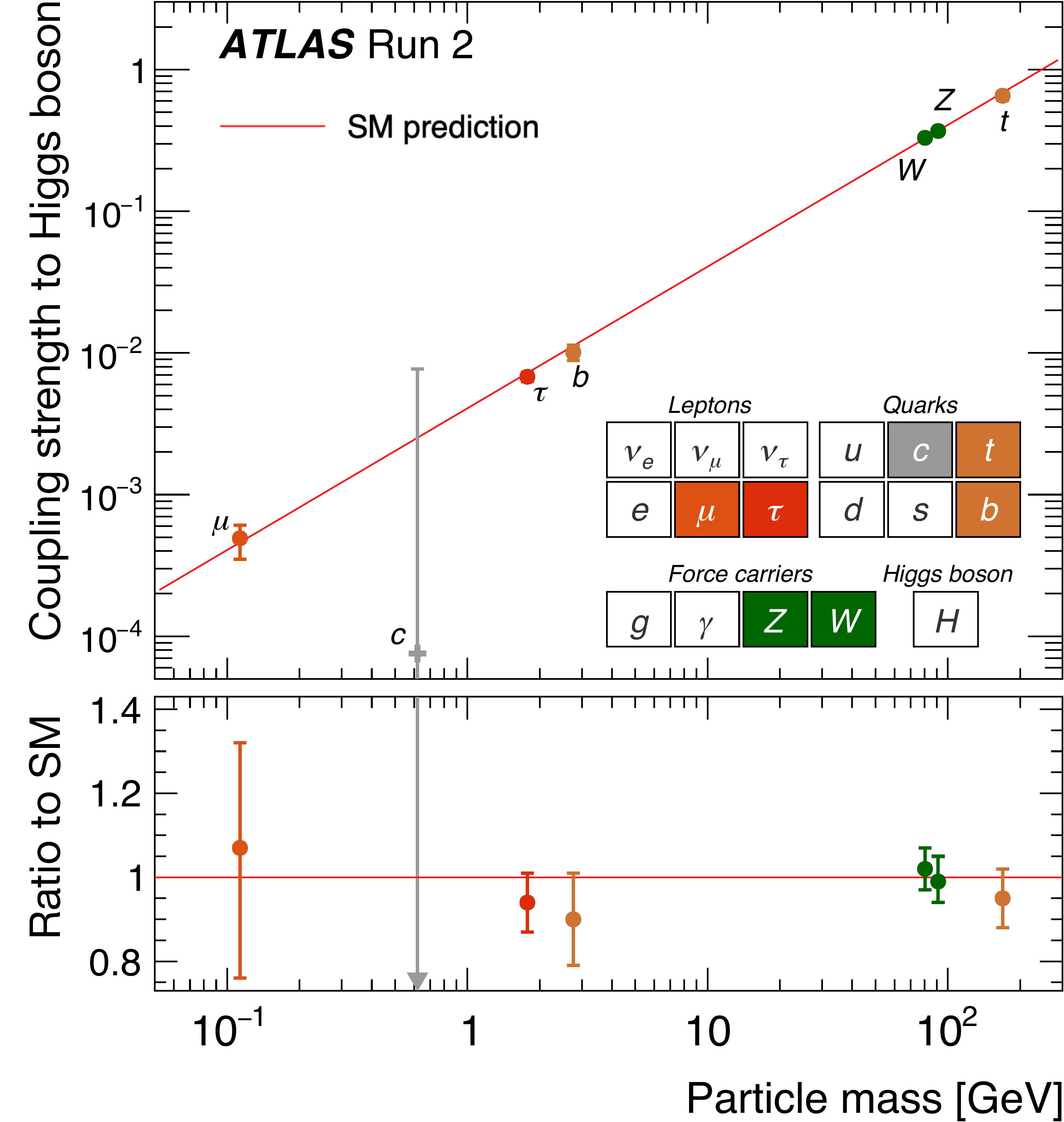}    
  \caption{{\bf Status of our knowledge of Higgs interactions with
      known particles}. Left: summary of which Higgs interactions have
    been conclusively established and future prospects.  Photons and
    gluons are omitted because they are massless and do not interact
    directly with the Higgs field. Neutrinos are also omitted: their
    masses are very small relative to those of the other leptons
    shown, and not individually known.
    Right: plot of measured strength of
    interaction of particles with the Higgs boson versus particle
    mass, as determined by the ATLAS Collaboration (adapted from
    Ref.~\cite{ATLASNature}). 
    The straight line shows the expected Standard Model (SM) behaviour,
    where the interaction strength is proportional to the mass of
    the fermions (squared mass for $W$ and $Z$ bosons).
    The CMS Collaboration has similar results~\cite{CMSNature}.
  } 
  \label{fig:interactions}
\end{figure}

There are several potential concerns about the robustness of these
kinds of test.
For instance, in the process shown in Fig.~\ref{fig:higgs-prod-decay}
there is an \emph{assumption} that there was a quantum fluctuation
producing a top anti-top pair.
Even if that assumption is correct, the number of events in the peak
tells us about the product of the top and $Z$ interactions, not the
top and $Z$ interactions separately.
For this reason, the LHC experiments look for the Higgs boson in a
multitude of production and decay processes, each one with
complementary sensitivity.
For example, it is possible to observe Higgs boson decays in events
where top-quarks are not simply an evanescent quantum fluctuation, but
are instead produced as short-lived real particles that emerge in
their own right from the collision together with the Higgs boson and
can be experimentally detected. 
Doing so~\cite{ATLAS:2018mme,CMS:2018uxb}, in 2018, was a major
milestone in particle physics, as were
the highly challenging observations of the Higgs boson decaying
  to bottom
quarks~\cite{ATLAS:2020fcp,CMS:2018nsn} and
$\tau$-leptons~\cite{ATLAS:2022yrq,CMS:2021gxc}.
Together, these measurements conclusively established that the Higgs
mechanism is responsible for the mass of the full third generation of
charged fermions.

Overall, by assembling information from different production and decay
channels, a picture has emerged of Higgs interactions for the heaviest
particles --- both vector bosons and fermions --- that is consistent
with the Standard Model hypothesis to within the current measurement
accuracies that range from $5\%$ to $20\%$, as summarised in
Fig.~\ref{fig:interactions}.
On the other hand, interactions with very light particles, such as the
electron and up and down quarks of which we are made of, are too rare for
current methods to observe.

While the discovery of the Standard Model Higgs boson was
highly anticipated
at the LHC, the ability to explore so many of its features was
a surprise.
To have established even part of the broad picture of Higgs-boson
interactions in just ten years is a major achievement, especially when
one considers that at the time when the LHC was being commissioned, many
of the production and decay channels that are central to
today's measurements were believed to be beyond the reach of the
LHC~\cite{CMS:2007sch,ATLAS:2009zsq}.

There are many reasons why this progress has been possible.
One of them is that nature happens to have chosen a value for
the Higgs mass that is particularly fortunate for experimental
studies.
Had the Higgs boson been $50 \GeV$ heavier, it would have been almost
impossible to detect more than just two basic decay channels (to a
pair of $W$ bosons or a pair of $Z$ bosons).
Had it been just $10\GeV$ lighter, the decays to $W$ bosons and $Z$ bosons
would probably have been impossible to see so far.
It was not just a question of good fortune, however.

The excellent performance of the LHC accelerator and of the ATLAS and
CMS detectors, each of them a highly complex system, has been crucial.
Furthermore, in the past ten years, there have been major advances in
techniques for analysing collider data.
One facet has been to learn how to reliably extract information about
individual proton--proton collisions when detectors contain not just
one proton--proton collision at a time, but dozens filling the
detector simultaneously, 40 million times per
second~\cite{Cacciari:2007fd,Bertolini:2014bba}.
Another reflects the fact that the beautifully clear peak in
Fig.~\ref{fig:higgs-prod-decay} (right) is the exception rather than
the rule:
for most other Higgs-boson studies ({\it e.g.}\ Higgs decay to
two bottom quarks or
two $W$ bosons), experimenters and theorists have
had to develop a wide range of technology for differentiating
Higgs-boson signals from the many processes with signatures
similar to that of a Higgs boson, but that do not involve a Higgs
boson.
These studies are increasingly benefiting from
a combination of new ideas for
how to perform the analyses ({\it e.g.}\ Ref.~\cite{Marzani:2019hun})
and the power of machine
learning~\cite{Guest:2018yhq}.

The quantitative interpretation of observed signal rates in terms of
Higgs interaction strengths would also not have been possible without
several decades' progress in the prediction and modeling of the rich
array of
effects that occur when protons collide, often associated with the
strong interaction.
It is crucial, for example, to have excellent theoretical control over
the rate of quark and gluon collisions given a certain number of
proton collisions~\cite{Gao:2017yyd,Ball:2022hsh}.
Another facet is that collisions often involve not just one quantum
fluctuation as in Fig.~\ref{fig:higgs-prod-decay}, but multiple
additional quantum fluctuations, each one of which modifies the
probability of Higgs boson production.
The greater the number of quantum fluctuations that one can account
for in theoretical predictions (today up to three additional
fluctuations~\cite{Anastasiou:2015vya}), the more accurately one can
relate experimental observations to the Standard
Model~\cite{Heinrich:2020ybq,LHCHiggsCrossSectionWorkingGroup:2016ypw}.
Finally, Fig.~\ref{fig:higgs-prod-decay} is a vastly
simplified picture and the experiments rely profoundly on
accurate simulation~\cite{Buckley:2011ms,Campbell:2022qmc} of the full
structure of proton--proton collisions, involving the production of
hundreds of particles per collision.
%

\section*{What is still to be established?}

In many respects, the experimental exploration of the Higgs sector is
only in its infancy.
There are two broad directions of ongoing investigation: obtaining
higher precision in studies of interactions that have already been
observed, and detecting further kinds of interactions that are, so
far, yet to be seen.

We start with the question of precision.
Examining Fig.~\ref{fig:interactions} (right), one sees that the
interactions of the Higgs boson with $W$ and $Z$ bosons and the third
generation charged leptons and quarks are currently known to a
precision of about $5$ to $20\%$.
We would not consider the theory of electromagnetism established if we had only
verified the strength of electromagnetic forces to within $10\%$
accuracy.

One of the reasons for aiming for higher precision is that while the
Standard Model Higgs mechanism outlined above is the simplest model
that is consistent with data, it is far from being the only viable
one.
Indeed, as we shall elaborate on below, it is widely believed that the
Standard Model as it stands cannot be a complete description of nature.
For example, it is conceivable that the Higgs boson is not an
elementary particle, but rather is composed of other, yet-to-be
discovered particles.
High-precision measurements of Higgs-related processes can be very
sensitive to such extensions of the Standard Model.
In particular, the rates of Higgs-related processes could be affected
by quantum fluctuations involving any new particles. Such effects
might be visible even in scenarios where the new particles are too
heavy to be directly produced and observed at a given collider.
In general, increasing the precision by a factor of four effectively
doubles the mass scale that can be indirectly probed for those new
particles.

The path for improvement is conceptually straightforward: with twenty
times more data to come in the next 15--20 years from the
approved high-luminosity upgrade of the LHC, and foreseeable improvements in
analysis techniques and theoretical calculations, the ATLAS and CMS
experiments expect to determine the currently observed set of
interactions to within a couple of percent~\cite{Dainese:2019rgk}.
Reaching beyond that requires a different kind of collider.
An electron--positron collider with centre-of-mass energies of around
$250\GeV$ (a ``Higgs factory'')~
\cite{FCC:2018evy,Roloff:2018dqu,Baer:2013cma,CEPCStudyGroup:2018ghi,Bai:2021rdg}
is widely considered to be a promising option (cf.\ the European Strategy
for Particle Physics~\cite{CERN-ESU-015}).
Advantages are that electrons and positrons, in contrast to protons,
are simple fundamental particles, and that the main Higgs-boson production
mechanisms at an electron-positron collider are largely free of
complications associated with strong interactions.
Such a collider could improve the precision of our knowledge of the
Higgs interactions by a further factor of about ten~\cite{deBlas:2019rxi}.

Let us now turn to discuss interactions that are yet to be observed.
Notwithstanding the good prospects for dramatically improving the
precision of Higgs measurements connected with the vector bosons and 3$^\text{rd}$
generation (heaviest) quarks and leptons, 
recall that the relevance of the Higgs sector for our everyday life is
that it is believed to generate masses for the 1$^\text{st}$
(lightest) generation of fundamental particles, the electron and up and
down quarks.
While experimentally testing our theoretical expectations for the interactions between 1$^\text{st}$ generation fermions and the Higgs boson is highly
challenging, there are prospects for the 2$^\text{nd}$ generation and
in particular the interactions of the Higgs boson with the muon, which
can be observed via the $H \to \mu^+\mu^-$ decay.
So far the data is suggestive of such
decays~\cite{ATLAS:2020fzp,CMS:2020xwi}, and definitive observation of
$H \to \mu^+\mu^-$, if it occurs at a rate that is compatible with the
Standard Model, is expected to come in the next decade.
Measurements involving the rest of the 2$^\text{nd}$ generation are more difficult.

The LHC can exclude anomalously large interactions of the Higgs boson
with charm quarks~\cite{Dainese:2019rgk} ({\it e.g.}\ using ideas such as
those in Refs.~\cite{Bishara:2016jga,Soreq:2016rae}).
It has long been thought that to definitively observe $H \to c\bar c$
decays would require a future $e^+e^-$ collider (or alternatively an
electron--proton collider~\cite{Andre:2022xeh}).
Significant recent improvements in sensitivity to this decay channel
at the LHC~\cite{ATLAS:2022ers,CMS:2022psv} raise the question of
whether future developments can bring its observation within reach of
the high-luminosity LHC.
For other Yukawa interactions, the path is less clear.

Investigations are ongoing to establish the potential sensitivity of a
future $e^+e^-$ collider to electron and strange-quark Yukawa
interactions (see {\it e.g.}\ Ref.~\cite{dEnterria:2021xij}), though
currently it seems that it will be challenging to obtain a
statistically conclusive signal.
For the up and down quarks' couplings to the Higgs boson, there are currently no concrete
possibilities in sight unless those couplings are very strongly enhanced
relative to the Standard Model
expectation. There has been
  discussion of whether precise atomic physics measurements could be
  sensitive to the Higgs forces involving light quarks~\cite{Delaunay:2016brc},
  however this seems challenging~\cite{Flambaum:2017onb}.

Central to all of Higgs physics is the Higgs potential.
Recall that the Higgs field is non-zero everywhere in the universe, and so produces
non-zero masses for fermions and electroweak bosons, because the
minimum of the Higgs potential, Eq.~(\ref{eq:Vphi}) and
Fig.~\ref{fig:higgs-potential}, lies at a non-zero value of the Higgs
field $\phi$.
One of the most important open questions in Higgs physics is whether
the potential written in that equation is the one chosen by nature.
We cannot directly explore the potential across different values of the
Higgs field.
However, it turns out that the specific shape of the potential in the
immediate vicinity of 
the minimum
determines the probability of an important process,
the splitting of a Higgs boson into two (or even
three) Higgs bosons --- this kind of process is referred to as a
Higgs-boson self interaction.
Accurate observation of such a process is widely considered to be the
best (but not the only~\cite{McCullough:2013rea}) way of
experimentally establishing whether the world we live in is consistent
with that simple potential.
By the end of the high-luminosity LHC's running in 15--20 years, the ATLAS and CMS
experiments are expected to see first indications of the simultaneous
production of two Higgs bosons.
However, gathering conclusive evidence for a contribution to Higgs-pair
production  from the splitting of a first Higgs boson almost
certainly requires a more powerful
collider and several options are under
discussion~\cite{FCC:2018vvp,CEPCStudyGroup:2018rmc,Roloff:2018dqu,Franceschini:2021aqd,Delahaye:2019omf}.

These are but some of the questions that are being explored.
Other important ones that the LHC experiments are starting to be
sensitive to include the lifetime of the Higgs
boson~\cite{Caola:2013yja,Campbell:2013una,ATLAS:2018jym,CMS:2022ley}
and the nature of Higgs interactions at energies well above the
electroweak energy scale~\cite{ATLAS:2020jwz,CMS:2020zge}.

\section*{Higgs and major open questions of particle physics and cosmology}

Many of the above measurements are of interest not just owing to the fundamental nature of the
Higgs sector within the Standard Model, but because they
are also sensitive to scenarios that extend the role of the Higgs
sector beyond that in the Standard Model.
While the Standard Model has successfully passed 
all the numerous experimental tests so far, it leaves open
several major questions.
To various degrees, the Higgs boson is tied to potential solutions
to these puzzles.

\begin{figure}[htb]
  \centering
  \includegraphics[width=0.9\textwidth]{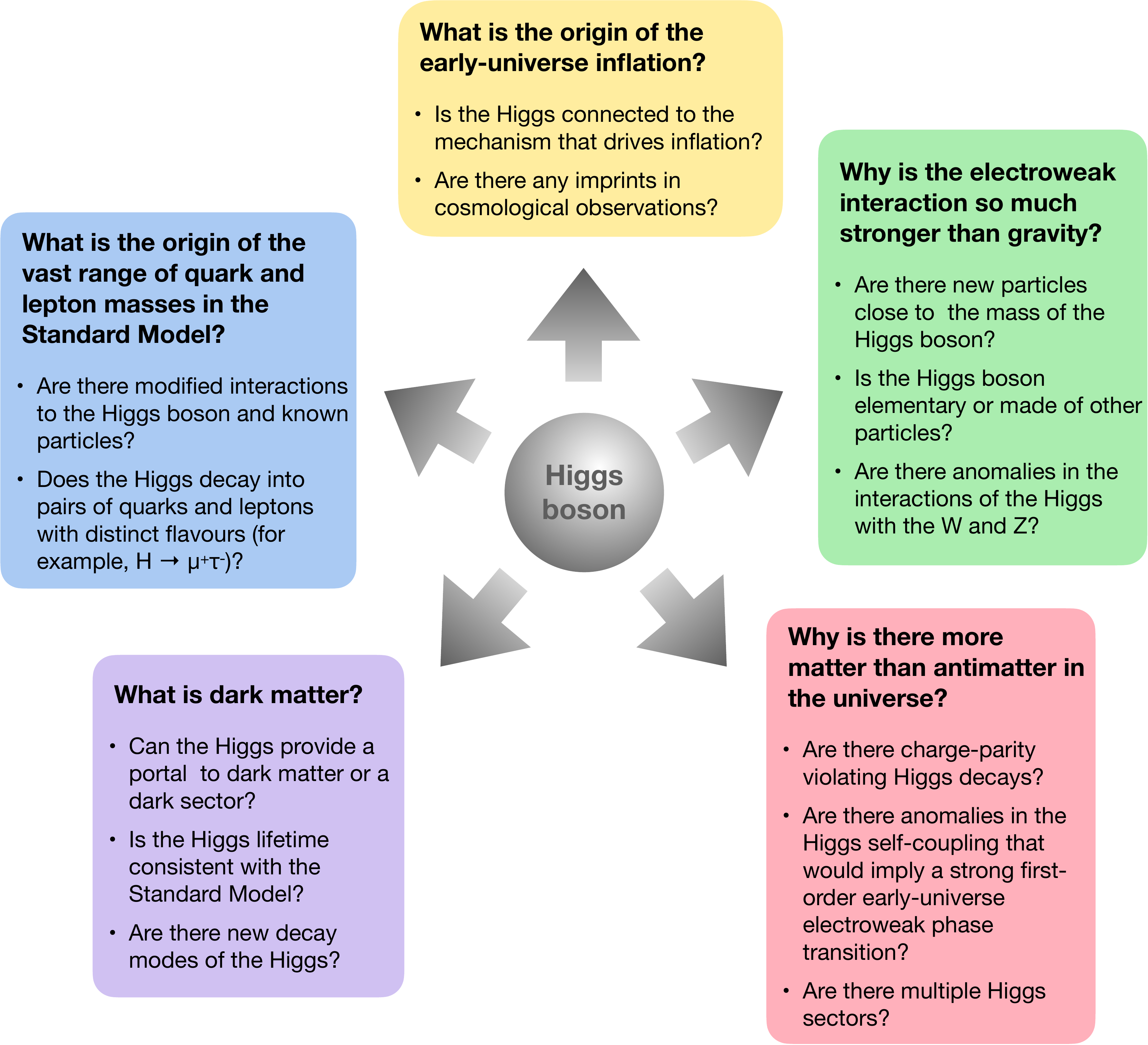}  
    \caption{{\bf Possible connections of Higgs physics with major open questions.}
      There are several major open questions in particle physics motivated by experimental observations or theoretical arguments. The Higgs boson could be the key to unravel some of these problems.
 }
 \label{fig:HiggsNP}
\end{figure}

We close our discussion with an overview of some of these
possible connections, illustrated in Fig.~\ref{fig:HiggsNP}, 
since they play an important role in guiding ongoing experimental and
theoretical research directions in particle physics.
There is a lot of ground to cover, so we will begin with and give more emphasis to aspects closely related to the Higgs boson, and only briefly mention later some of the more speculative ideas.

One major puzzle is that the weak and Higgs interactions are
much stronger, by a factor of about $10^{32}$, than the gravitational
interaction.
This is especially challenging if one harbours the hope --- as do many
physicists --- that all the known interactions might come from a
unifying and simpler framework.
Over the past 
decades, the desire to explain the origin of
this large difference, the so called ``hierarchy problem'', has
motivated a range of theoretical proposals.

One possibility is for the Higgs boson not to be an elementary particle, but rather a composite object made of other, as yet undiscovered
particles~\cite{Kaplan:1983fs}.
Examples of other well studied proposals are
new (approximate) space-time
symmetries~\cite{Fayet:1976et,Fayet:1977yc,Dimopoulos:1981zb}, and new
space
dimensions~\cite{Arkani-Hamed:1998jmv,Antoniadis:1998ig,Randall:1999ee,Randall:1999vf}. More recently, some more speculative ideas suggested possible connections between the weak scale and cosmological
evolution~\cite{Graham:2015cka,Arkani-Hamed:2016rle,Giudice:2021viw} or the amount of
dark energy in the
universe~\cite{Arvanitaki:2016xds,Arkani-Hamed:2020yna}.

Without one of these proposals, or a new mechanism yet to be thought of, 
the hierarchy between the weak and the gravitational interaction can only arise if distinct parameters in some
ultimate fundamental theory cancel to within $1$ part in $10^{32}$. 
This is known as the fine-tuning problem of the Higgs sector.

The discovery of the Higgs boson brought such questions unavoidably to
the fore. The mere existence of the Higgs boson, and the (still approximate) picture of its properties, already exclude many theoretical ideas. In comparison with the decades before its discovery, we now have much clearer target and sharper questions to answer with our theoretical models.

Another important question is why there is more matter than
anti-matter in the universe.
This so called baryonic asymmetry cannot be explained within the Standard
Model.
Such an asymmetry can be generated if a suitable set of conditions is
met~\cite{Sakharov:1967dj}.
One promising avenue that is being explored follows the history of the
universe as it cooled down after the big bang.

When the universe was very hot, the minimum of the Higgs potential at
a non-zero value of the Higgs field was largely irrelevant because
temperature fluctuations were much larger than the depth of the
potential.
As the universe cooled, the situation changed.
Within the Standard Model that change is smooth. Other promising scenarios, which involve new particles interacting with the Higgs boson, would
generate a sharper  transition, which sets the
stage for generating the observed baryon asymmetry~\cite{Cohen:1993nk}, though
further ingredients are also needed.

These scenarios involve more complex structures for the Higgs
potential, and at least one new particle at the electroweak energy scale, which can be
searched for at the LHC either through its direct production or
through its indirect impact on the Higgs couplings, in particular the
Higgs self interaction. 
A measurement of the latter is therefore essential to shed light on
this question.
Early-universe phase transitions could also produce gravitational signatures that 
can be detected by future gravitational wave
experiments~\cite{Caprini:2015zlo,Caprini:2019egz}. 

In addition to the questions directly related to the Higgs boson
mentioned above, there are also other contexts in which the Higgs boson can play an important role.  One example of this is the question of dark matter. 
Astrophysical and cosmological observations show 
that the majority of the matter in the universe is dark and not made
of any particle we know of.
Such observations rely on the
gravitational effects of the dark matter on ordinary, Standard Model
matter.
At the same time, we know very little about the non-gravitational
properties of dark matter.
New particles with masses around
the electroweak and Higgs mass scales can be promising dark matter
candidates.

Since the Higgs mechanism is responsible for generating similar masses
of the Standard Model particles, it is possible that it plays some role
in generating the dark matter mass as
well
\cite{Silveira:1985rk,Burgess:2000yq,McDonald:1993ex}.
There are also scenarios where the dark matter sector involves more
than one kind of particle.
Similar to particles in the Standard Model, they could have their own
interactions, and a whole set of other closely related particles.
In this case, the Higgs boson would provide a portal to a new ``dark
world'' \cite{Patt:2006fw}.

The origin of the pattern of masses and interactions among different
generations of the Standard Model particles is an intriguing puzzle.
For example, first generation quarks are much
lighter than the third generation quarks, which in the Standard Model
needs to be arranged manually by setting correspondingly disparate
values of the Yukawa couplings.
Understanding the origin of this pattern has also
been the focus of decades of efforts.
Since the Higgs sector is responsible for generating the masses of
these particles, it is tempting to think that the actual Higgs sector
may be structurally different from the Standard Model, in a way that
causes the observed pattern to emerge naturally \cite{Barr:1976bk,Bjorken:1977vt,Babu:1999me}.

The models that explore such ideas often lead to predictions of
modified interactions between the Higgs boson and the quarks (and/or
leptons). One signature of such models is that the Higgs boson could
decay into a pair of quarks or leptons with different flavour.
Similarly, one may also ask whether the Higgs mechanism has a role in generating
the extremely small masses for neutrinos and various models have been
envisaged in this respect \cite{deGouvea:2016qpx}.

The questions above relate the Higgs boson with known or unknown
elementary particles.
However, there are also mysteries in fundamental physics that go beyond
such types of questions and speculative, yet intriguing links have been
proposed with the Higgs sector.
For example, it has been noted that 
the Standard-Model self interaction of the Higgs boson becomes very
close to zero if it is measured~\cite{Cabibbo:1979ay,Hung:1979dn,Lindner:1985uk} 
at energies nine orders of magnitude beyond the Higgs
mass~\cite{Degrassi:2012ry,Buttazzo:2013uya}. 
A curious and connected fact 
is that it seems likely that the Standard-Model Higgs
sector has a ground state with lower energy than the state we live in.
Hence, quantum mechanics would allow a ``tunneling'' process through
which our whole universe can decay, even though the probability of
such an event happening within the 14-billion year age of the universe
is tiny.
The final possibility that we mention for new dynamics of the Higgs
field at high energies is a possible link to inflation, which is a
period of exponential expansion in the early universe that is
essential to explain the striking long-distance uniformity of the
cosmic microwave background. The Higgs boson, having spin-0,
may be responsible to drive inflation~\cite{Bezrukov:2007ep}.

The Higgs boson is an invaluable tool in the search for answers to
several of the above questions.
Many of the proposed solutions predict the existence of new particles
that generally interact directly with the Higgs boson. These particles
are actively searched for at high-energy colliders. 
Still, even if the direct production of these particles lies outside our reach, for instance because the LHC is not energetic enough, their
involvement in quantum fluctuations may affect Higgs boson
production and decay, in the same way that top-quark quantum fluctuations
mediate Higgs production in Fig.~\ref{fig:higgs-prod-decay}.
The expected future advances in precision measurements of the Higgs
boson, as mentioned above, will bring considerably improved
sensitivity to such scenarios.

\section*{Conclusions}

The discovery of the Higgs boson at the LHC marked the beginning of a
new era of particle physics.
In the ten years since, the exploration of the Higgs sector has
progressed far beyond original expectations, thanks to ingenious
advances both experimental and theoretical.
Every Higgs-related measurement so far has been consistent with the
Standard Model, the simplest of all current models of particle
physics: a remarkable win for Occam's razor. 
Today, it is clear that the Higgs mechanism, first proposed in the
1960's, is responsible not only for the masses of the $W$ and $Z$
bosons and but also for those of the three heaviest fermions.
This directly implies the existence of a fifth force, mediated by the
Higgs boson.
Still, much remains to be probed.
Whatever is found in the coming decades, we will be wiser: either with
solid evidence for parts of the Standard Model that remain crucially
to be established, such as the nature of Higgs potential, or opening a
window to new horizons and major mysteries of the universe.

\section*{Acknowledgments}

Our perspective on the Higgs sector has been shaped by exchanges
with many of our colleagues.
In this regard we especially acknowledge conversations with
Nima Arkani-Hamed,    
Fabrizio Caola,       
Gian Giudice,         
Christophe Grojean,   
Ulrich Haisch,        
Marumi Kado,
Greg Landsberg,       
Michelangelo Mangano, 
Matthew McCullough,   
Paolo Nason,          
Matthew Reece,        
G\'eraldine Servant,  
Raman Sundrum,        
Carlos Wagner         
and thank a number of them also for valuable comments on this
manuscript.
GPS is supported by a Royal Society Research Professorship
(RP$\backslash$R1$\backslash$180112) and by the United Kingdom's
Science and Technology Facilities Council (grant ST/T000864/1).
LTW is supported by the US Department of Energy grant DE-SC0009924.

\bibliography{higgs}{}
\bibliographystyle{JHEP}

\end{document}